\documentclass{article}
\usepackage{spconf,amsmath,graphicx,hyperref}
\usepackage{booktabs}
\usepackage{pifont}
\usepackage{xcolor}

\definecolor{Ctime} {RGB}{220,220,220}
\definecolor{Ctype} {RGB}{205,175,175}
\definecolor{Ccoord}{RGB}{175,195,215}
\definecolor{Ccurve}{RGB}{195,210,180}
\definecolor{Cbeats}{RGB}{215,200,170}
\definecolor{Ccol}  {RGB}{195,185,210}
\definecolor{Cspec} {RGB}{45, 45, 45}
\definecolor{Cdon}  {RGB}{220,165,160}
\definecolor{Ckat}  {RGB}{160,190,215}

\newcommand{\tok} [2][white]{\colorbox{#1}{\strut\scriptsize\textsf{#2}}}
\newcommand{\ttime}[1]{\tok[Ctime]{#1}}
\newcommand{\ttype}[1]{\tok[Ctype]{#1}}
\newcommand{\tcoord}[1]{\tok[Ccoord]{#1}}
\newcommand{\tcurve}[1]{\tok[Ccurve]{#1}}
\newcommand{\tbeats}[1]{\tok[Cbeats]{#1}}
\newcommand{\tcol}  [1]{\tok[Ccol]{#1}}
\newcommand{\tspec} [1]{\tok[Cspec]{\textcolor{white}{#1}}}
\newcommand{\tdon}  [1]{\tok[Cdon]{#1}}
\newcommand{\tkat}  [1]{\tok[Ckat]{#1}}

\title{Transfer Learning for Rhythm Game Chart Generation  \\ from Music Transcription and Across Games}
\name{Sojeong An$^{1}$ \qquad Nathan Pruyne$^{2}$ \qquad Chris Donahue$^{2}$ \qquad Dasaem Jeong$^{1}$}
\address{$^{1}$Music \& Arts Learning (MALer) Lab, Sogang University, South Korea \\
         $^{2}$Computer Science Department, Carnegie Mellon University, Pittsburgh, USA}

\begin{document}
\ninept
\maketitle
\begin{abstract}
Rhythm games ask the player to act in time with a recording, following a chart: a sequence of actions, each with a time and, depending on the game, a lane, type, or playfield position. Charts are hand-written per recording and difficulty. Automatic generation from audio has been studied with pipeline and end-to-end approaches, and recent work reuses pre-trained audio encoders, but never compares them with one trained from scratch. None of these targets automatic music transcription (AMT), yet most chart times are note onsets, which AMT models locate. We freeze a pre-trained audio encoder, an AMT model in particular, and train only a linear projection and a Transformer decoder on charts. With MuScriptor, this fits human charts more closely than encoders trained from scratch and places more accurate note types than MERT, a music encoder pre-trained without transcription labels. Joint training on three osu! modes raises event F1 on the mode with the fewest charts from 0.492 to 0.506, while charts from \textit{Clone Hero}, similar in layout, lower it to 0.482. Transfer between games thus appears to depend on charting style and the music covered.
\end{abstract}

\section{Introduction}
\label{sec:intro}

\begin{figure}[!t]
\centering
\includegraphics[trim=18 65 18 18, clip, width=0.75\columnwidth]{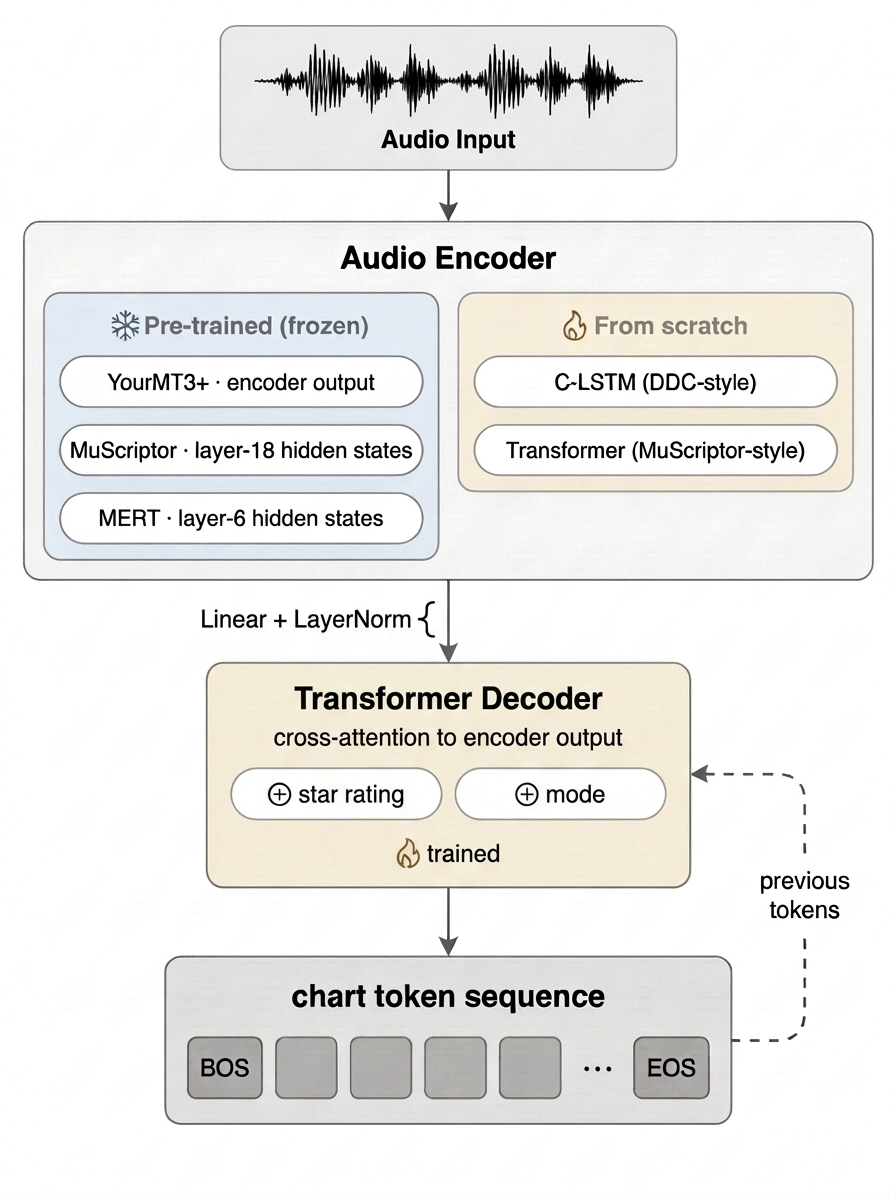}
\caption{Overview of the system.}
\label{fig:system}
\end{figure}

Rhythm games require the player to perform actions in time with a recording: marks scroll across the screen, and as each one arrives the player presses a key, strikes a drum, or clicks a point on a playfield. These marks constitute a \textit{chart}, a sequence of actions that each carry a time and, depending on the game, a lane, a type, or a two-dimensional position. Charts are authored by hand, one for each recording and difficulty. A chart is tightly conditioned on its audio, in that its times align with onsets a listener can hear, but unlike a transcription it is an interpretation: the author selects which sounds to mark, what to emphasize, and how the difficulty should develop. The audio therefore does not determine a single correct chart, and many charts are valid for one recording, which places the task between transcription and symbolic music generation.

Automatic chart generation from audio has been studied both with pipelines that detect note onsets and then assign a type to each~\cite{ddc, genelive,tcp2025}, and with end-to-end models~\cite{yi2023}. Recent work takes the audio representation from a model pre-trained on another task, such as audio compression~\cite{audio2chart} or speech recognition and general music understanding~\cite{zhang2026}, but does not compare it with an encoder trained from scratch, so how much the pre-training helps is not known. Nor is any of these tasks music transcription. Although many charts are valid for one recording, most times in each of them are note onsets, which automatic music transcription (AMT) models are trained to locate. To the best of our knowledge, an AMT encoder has never been used for chart generation.

We study transfer learning for chart generation along two dimensions. The first is transfer from a pre-trained audio encoder, an AMT model in particular. We freeze a pre-trained encoder and train only a linear projection and a Transformer decoder on charts (Fig.~\ref{fig:system}).
With MuScriptor~\cite{muscriptor} as the frozen encoder, event F1 rises from 0.323 to 0.433 over the same architecture trained from scratch. MERT~\cite{mert}, a music model pre-trained without transcription labels, also outperforms the encoder trained from scratch and is comparable to MuScriptor on the onsets, but places a less accurate object type at each one. Of the two AMT models, MuScriptor also gives a stronger chart model than YourMT3+~\cite{yourmt3}.

\begin{figure*}[t!]
\centering
\setlength{\fboxsep}{1.5pt}
\newcommand{\panelcap}[1]{\\[2pt]{\small\textit{#1}}\\[3pt]}
\newcommand{\panelseq}[1]{\begin{minipage}[t]{\linewidth}\raggedright
\setlength{\parindent}{0pt}\spaceskip=1.2pt plus 0.6pt minus 0.3pt\relax #1
\end{minipage}}
\begin{minipage}[t]{0.24\textwidth}\centering
\includegraphics[width=\linewidth, trim=0 0 0 33, clip]{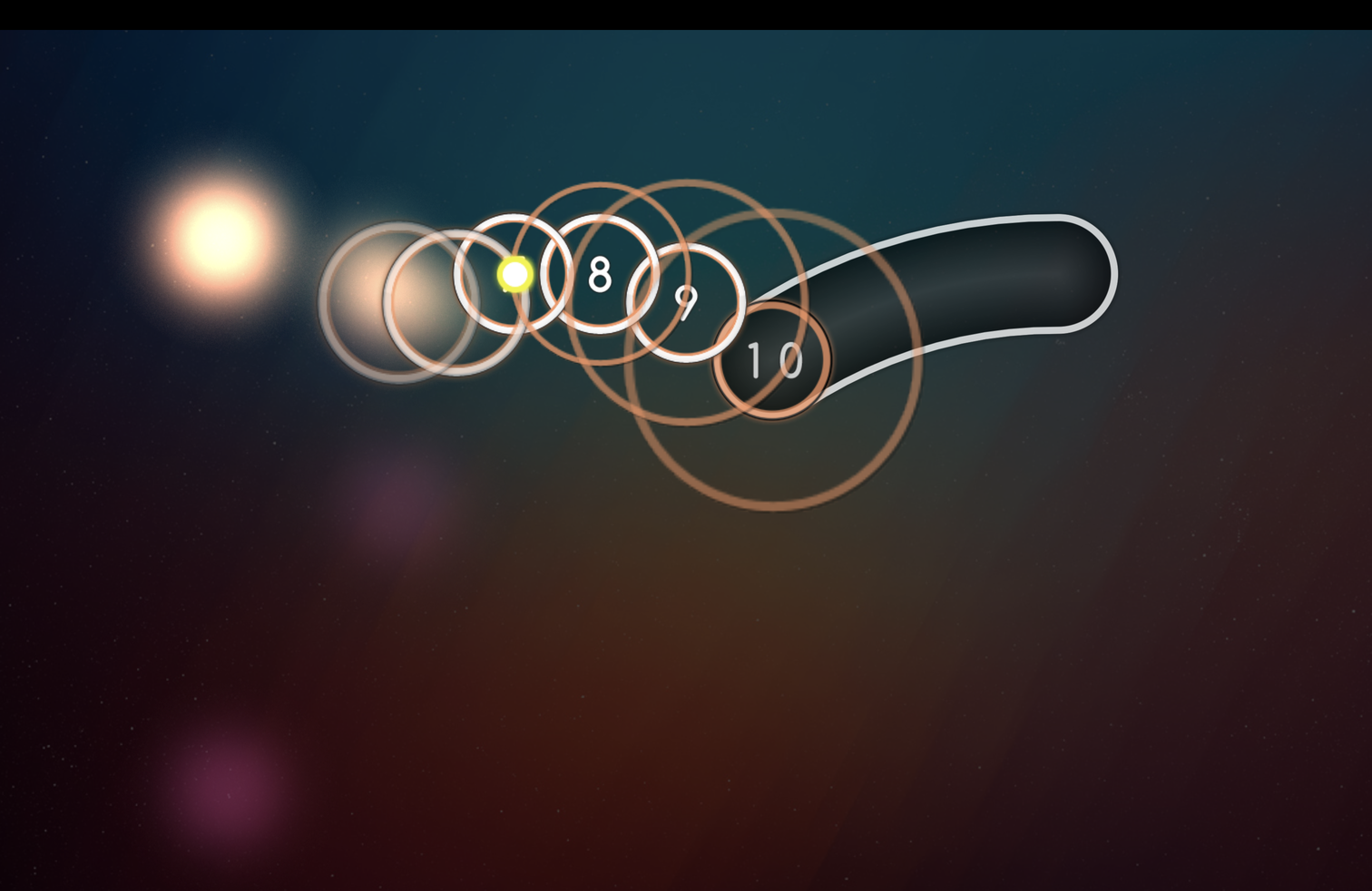}
\panelcap{standard}
\panelseq{\tspec{BOS} \ttime{t=12} \ttype{circle} \tcoord{x=18} \tcoord{y=7}~%
\ttime{t=30} \ttype{slider} \tcurve{B} \tcoord{x=5} \tcoord{y=9}~%
\tcoord{x=11} \tcoord{y=14} \tcoord{x=20} \tcoord{y=6}~%
\tbeats{beats=1.0} \tbeats{rep=1} $\cdots$ \tspec{EOS}}
\end{minipage}\hfill
\begin{minipage}[t]{0.24\textwidth}\centering
\includegraphics[width=\linewidth, trim=0 0 0 33, clip]{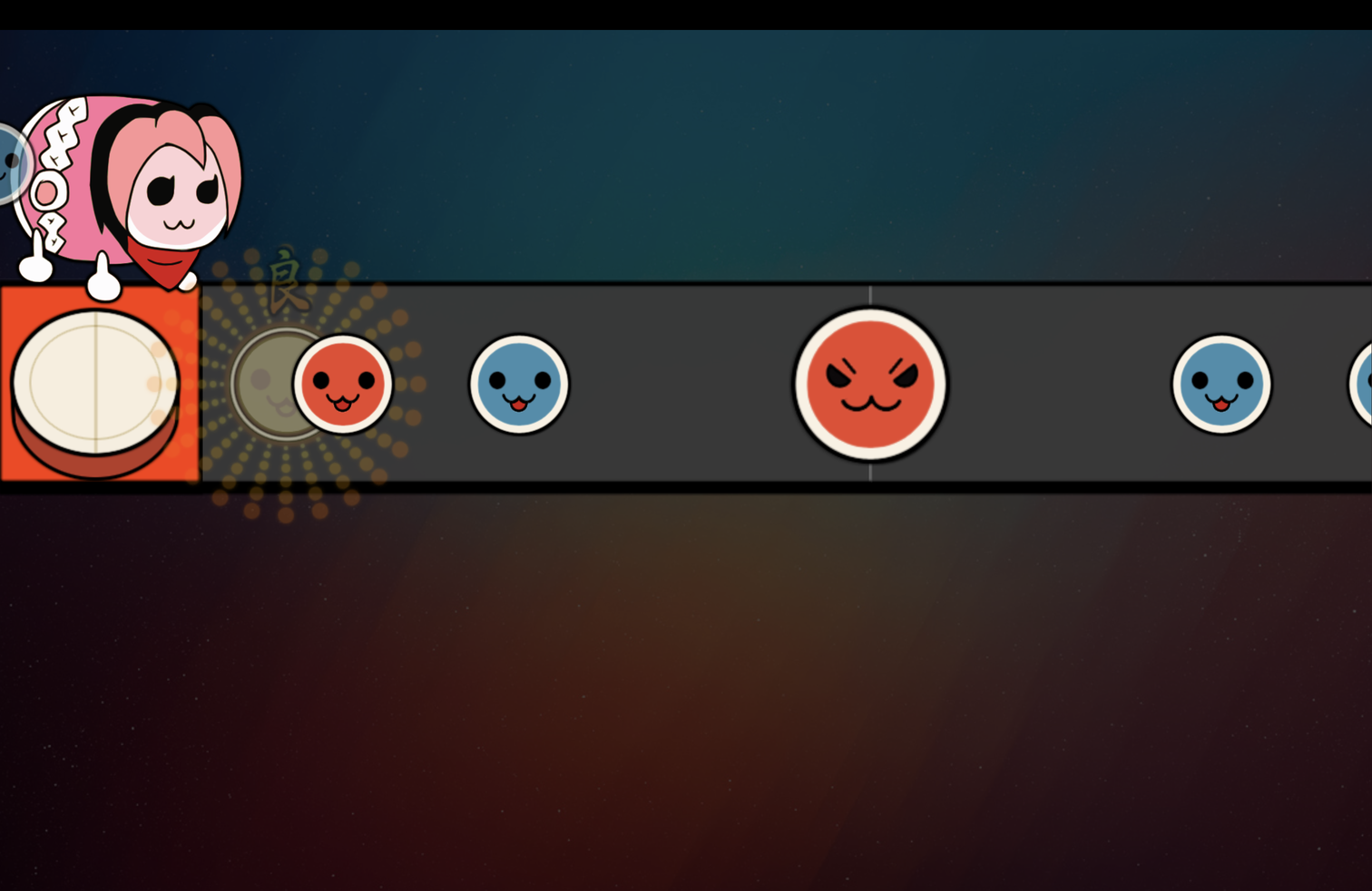}
\panelcap{taiko}
\panelseq{\tspec{BOS} \ttime{t=7} \tdon{type=0} \ttime{t=46} \tkat{type=1}~%
\ttime{t=80} \tdon{type=2} \ttime{t=95} \tkat{type=3} $\cdots$ \tspec{EOS}}
\end{minipage}\hfill
\begin{minipage}[t]{0.24\textwidth}\centering
\includegraphics[width=\linewidth, trim=0 0 0 33, clip]{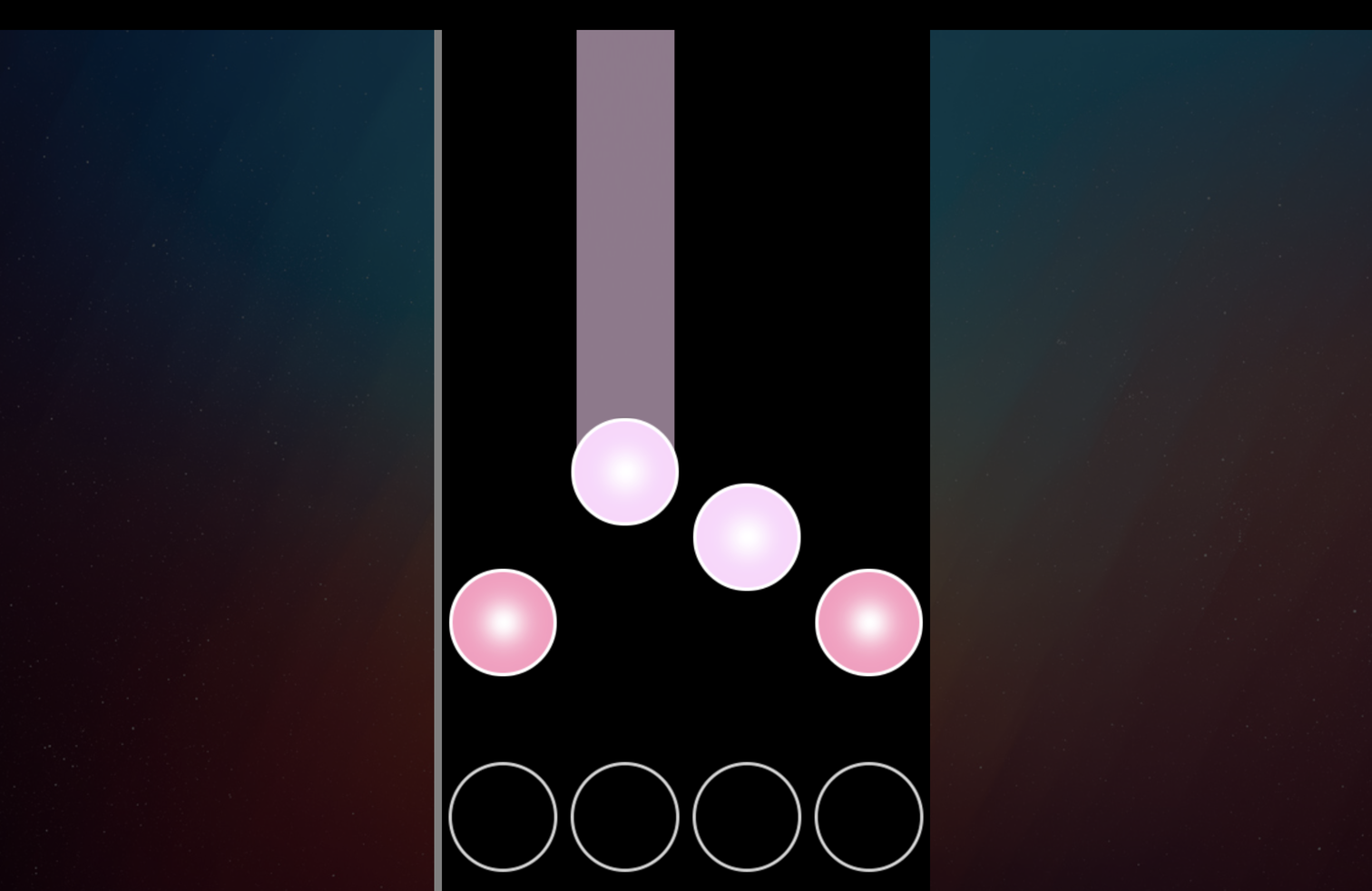}
\panelcap{mania}
\panelseq{\tcol{held} \tspec{BOS} \ttime{t=21} \tcol{col=0} \ttype{tap}~%
\ttime{t=29} \tcol{col=2} \ttype{hold$\downarrow$} \ttime{t=44} \tcol{col=2}~%
\ttype{hold$\uparrow$} \ttime{t=49} \tcol{col=1} \ttype{tap} $\cdots$ \tspec{EOS}}
\end{minipage}\hfill
\begin{minipage}[t]{0.24\textwidth}\centering
\includegraphics[width=\linewidth, trim=0 0 0 33, clip]{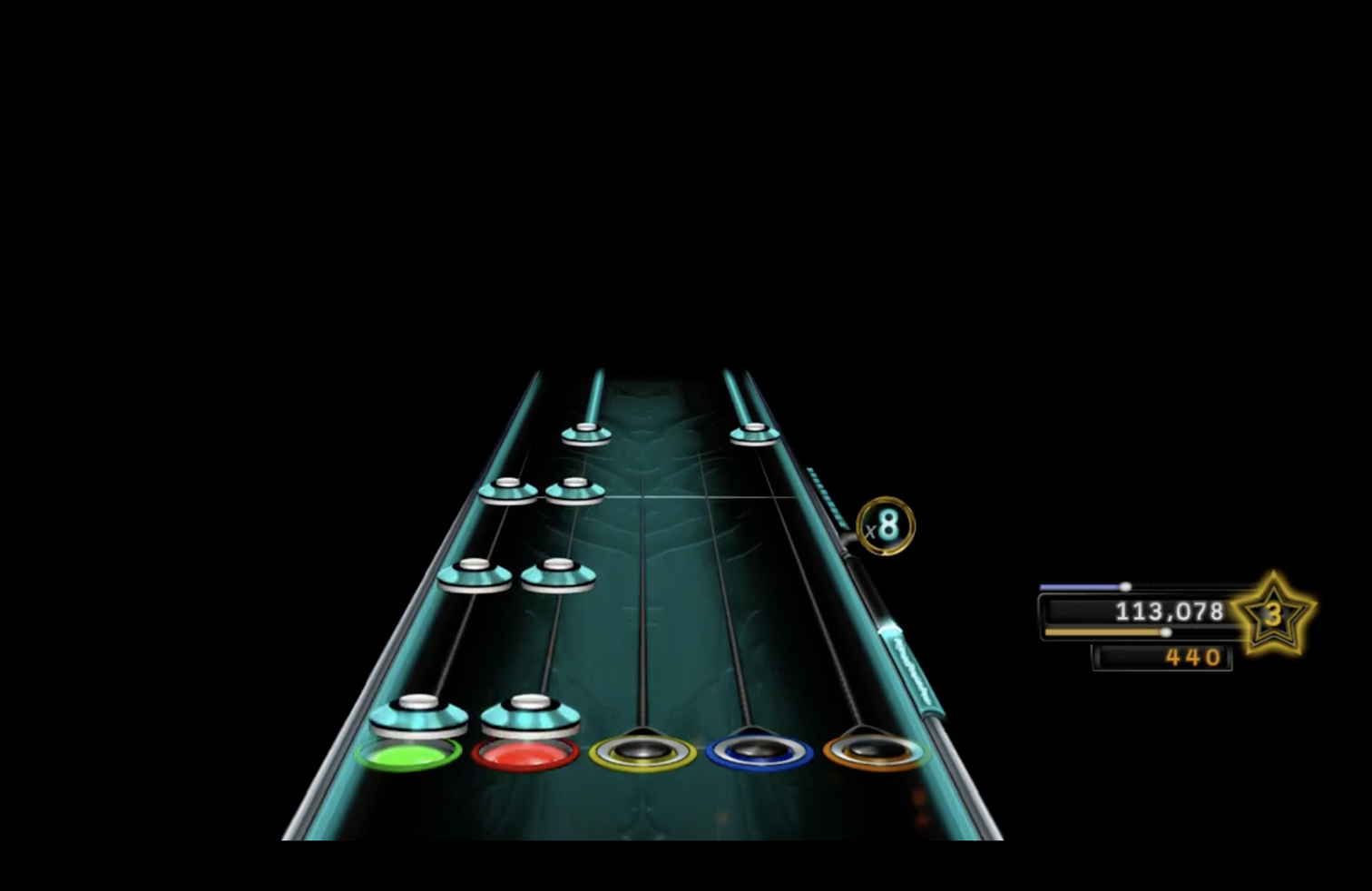}
\panelcap{Clone Hero}
\panelseq{\tspec{BOS} \ttime{t=12} \tcol{lane=3} \ttype{type=0}~%
\ttime{t=30} \tcol{lane=1} \ttype{type=1} $\cdots$ \tspec{EOS}}
\end{minipage}
\caption{Generated charts in play, each with the token sequence of one segment; the \textit{Clone Hero} panel shows a chart from the corpus. \textit{Standard} places notes freely on a 2D playfield, while \textit{taiko}, \textit{mania} and \textit{Clone Hero} assign each note a lane or a column. \textit{Mania} prepends the set of columns already held before BOS.}
\label{fig:modes}
\end{figure*}

The second is transfer across rhythm games and game modes. We train a single model on \textit{standard}, \textit{taiko} and \textit{mania}, three modes of the rhythm game osu! (Fig.~\ref{fig:modes}). In \textit{standard} notes are placed freely on a two-dimensional playfield, in \textit{taiko} every note is a drum hit on a single lane, and in \textit{mania} every note is a tap or a hold in one of four columns. Joint training raises event F1 on \textit{mania}, the mode with the fewest charts, from 0.492 to 0.506. We then pair \textit{mania} with a game outside osu!. \textit{Clone Hero} is a guitar game that is lane-based like \textit{mania}, but its charts follow the melody of a single instrument part, most often the guitar. Despite superficial similarities between the two games, transfer learning does not appear to provide benefits: training the two together lowers event F1 to 0.482.

\section{Related work}
\label{sec:rel}

Automatic chart generation from audio began as a pipeline that first locates note times and then assigns a type to each: the C-LSTM of Dance Dance Convolution (DDC)~\cite{ddc}, and for osu! \textit{mania} a support vector machine detecting onsets followed by rule-based column placement~\cite{perkasa2015} and a C-BLSTM trained on fuzzy onset labels~\cite{liang2019}. Later work improved one stage at a time, with beat conditioning and multiple temporal scales for the onset stage~\cite{genelive}, and for the type stage Time-based Chart Partitioning (TCP)~\cite{tcp2025}, which matches each segment's rhythm to a human chart and copies its types. In these pipelines, from DDC to TCP, only the onset stage hears the audio; the type stage sees the note times and rhythm features~\cite{ddc,liang2019,tcp2025}.
This is a reasonable simplifying assumption for certain rhythm games, but for others, the event types are more dependent on the audio.
In osu! for example,
a low drum hit calls for a \textit{don} and a high one for a \textit{kat} in \textit{taiko}, and a sustained note for a hold in \textit{mania}. TaikoNation~\cite{taikonation} and AutoOsu~\cite{autoosu} instead predict times and types jointly in one network, and Yi et al.~\cite{yi2023} and Zhang et al.~\cite{zhang2026} write the chart as a token sequence. Most of these models take a difficulty label as input~\cite{ddc,liang2019,autoosu,yi2023,zhang2026}, as we do with the star rating.

Each of these systems is trained and evaluated on one game, with its own metric and tolerance, so their scores cannot be compared with one another; TCP, for one, rules out DDC and TaikoNation as baselines because they were built for other games~\cite{tcp2025}. The one transfer result, Yi et al. pre-training on osu! \textit{mania} and fine-tuning on two StepMania datasets, is between games with the same four keys. Mapperatorinator~\cite{mapperatorinator}, a community system, trains one Whisper-architecture model on all four osu! modes, but no evaluation of it has been published. Whether what a model learns on one game carries to a game with a different playfield has therefore not been measured.

Two recent models take the audio representation from a network pre-trained on another task, but neither compares it with an encoder trained from scratch. audio2chart~\cite{audio2chart} keeps a neural codec frozen and compares it only with a model that hears no audio at all, and Zhang et al.~\cite{zhang2026} fine-tune Whisper or MERT together with the decoder and find Whisper the stronger, a comparison in which the architecture and the pre-training task change together and neither encoder is tested frozen or from scratch.
Neither task is music transcription; to the best of our knowledge, no work has taken the internal representation of a pre-trained AMT model as an audio encoder for another task, chart generation included.
Yet a frozen representation with a shallow head trained on top has carried over to MIR tasks from generative audio models~\cite{castellon2021} and from self-supervised encoders~\cite{mert,marble}, and the training objective of AMT~\cite{benetos2019amt} includes locating note onsets~\cite{onsetsframes,mt3}, the quantity most of a chart's timestamps carry.

{\sloppy We therefore take our encoders from two AMT models, YourMT3+~\cite{yourmt3} and MuScriptor~\cite{muscriptor} (Sec.~\ref{ssec:amt}).
We hold everything else fixed, so that an encoder trained from scratch, a frozen AMT encoder and a frozen self-supervised encoder are scored with the same decoder, data and metrics. The same setup then measures transfer across three osu! modes and to \textit{Clone Hero}.\par}

\section{Chart representation}
\label{sec:chart}

We write every chart as a sequence of timestamped events $(t_1,e_1),\allowbreak \dots,\allowbreak (t_N,e_N)$, and the three osu! modes and \textit{Clone Hero} differ only in what $e_i$ contains.
We cut each song into 5\,s segments, the input window of MuScriptor (Sec.~\ref{ssec:amt}), and write the chart of each segment as one token sequence (Fig.~\ref{fig:modes}). The segments do not overlap, and the decoder generates the chart of each segment independently of the others, from its audio alone; the charts are then concatenated.
Within a segment, a timestamp indexes a grid of 500 values about 10\,ms apart from the segment start. \textit{Taiko} encodes each note as a timestamp and one of six types. \textit{Mania} encodes a timestamp, one of four columns and one of three types: a tap or the start or end of a hold. \textit{Clone Hero} encodes a timestamp, one of five frets or an open strum, and one of four types, one of which ends a sustained note.
\textit{Standard} quantizes positions on its 512\,$\times$\,384 playfield to a 16-unit grid, 32 values in $x$ and 24 in $y$: a hit circle adds its coordinates; a slider adds a curve type, two to four anchors, a length in beats and a repeat count; a spinner adds no coordinates, since it always sits at the center of the playfield.
In \textit{mania} a hold can span a segment boundary, so we prepend one token giving the set of columns held at the start of the segment.
The resulting vocabularies hold 509 tokens for \textit{taiko}, 513 for \textit{Clone Hero}, 526 for \textit{mania} and 577 for \textit{standard}, of which 500 are timestamps in every case. The joint models take the union of their modes' vocabularies, 606 tokens for the three osu! modes and 536 for \textit{mania} with \textit{Clone Hero}, and share the timestamps across modes.

\section{Model}
\label{sec:model}

\subsection{Pre-trained encoders}
\label{ssec:amt}

MuScriptor~\cite{muscriptor} is a decoder-only Transformer that takes a 5\,s mel spectrogram as a prefix; we take the hidden states at layer 19 of 24, chosen with a linear onset probe, giving 501 frames of 1024 dimensions at 100\,Hz. YourMT3+~\cite{yourmt3} uses a PerceiverTF encoder that maps a 2.048\,s spectrogram to 110 frames of 24 latents in 128 dimensions, which we flatten to 3072 per frame; its window is shorter than our segment, so we slide it over the song in steps of 55 frames (1.03\,s) and keep the middle 55 of each, giving 267 frames of 3072 dimensions at 53\,Hz. MERT~\cite{mert} is a self-supervised encoder that takes the 24\,kHz waveform directly; we take the hidden states at layer 6 of 24, chosen with the same probe, giving 374 frames of 1024 dimensions at 75\,Hz. We take hidden states rather than the transcription the two AMT models output, so that the decoder sees the representation itself instead of a discrete note list.

We freeze all three encoders and train only a linear projection to the decoder dimension and a LayerNorm between them and the decoder. In preliminary experiments, fine-tuning the entire encoder led to overfitting on our data.

\subsection{Encoders trained from scratch}
\label{ssec:scratch}

We train two encoders from scratch as baselines. The first is the C-LSTM of DDC~\cite{ddc}: two convolutional layers of 10 and 20 filters and two LSTM layers of 200 units, taking a three-channel log-mel spectrogram (512, 1024 and 2048 FFT sizes, 80 mel bins, 10\,ms hop), with a linear projection to the decoder. A wider variant, grown from DDC's 0.7M parameters to 11.4M roughly in proportion to our larger dataset, scored no better on validation. The second isolates the effect of pre-training: a Transformer with MuScriptor's front end and architecture but random initialization, taking a single-channel log-mel spectrogram (2048 FFT size, 512 mel bins, 10\,ms hop) into 19 causal layers of width 1024, matching MuScriptor up to layer 19.

\subsection{Decoder and difficulty conditioning}
\label{ssec:decoder}

The decoder is a stack of standard Transformer decoder layers, with cross-attention to the audio representation and causal self-attention, and it predicts the next token.
For single-mode experiments we use six layers at $d_\text{model}=512$ for \textit{standard} and four layers at $d_\text{model}=256$ for \textit{taiko}. The joint model has eight layers at $d_\text{model}=768$, as do the \textit{mania}-only model and the model trained on \textit{mania} with \textit{Clone Hero}, so that only the training data differs among them.

We condition the decoder on the chart's \textit{star rating} --- osu!'s scalar measure of difficulty, where higher means harder. A two-layer MLP maps the star rating to a $d_\text{model}$-dimensional vector, and we add it to every token embedding before a LayerNorm; the MLP's last layer is zero-initialized so conditioning has no effect at the start of training. The joint models also add a zero-initialized mode embedding to every token embedding.

\section{Experiments}
\label{sec:exp}

\subsection{Data}
\label{ssec:data}

We take osu! beatmapsets \textit{ranked} --- written by players and approved through osu!'s ranking process --- between May 2019 and December 2025, and keep their \textit{standard}, \textit{taiko} and four-column \textit{mania} charts.\footnote{Retrieved from \url{https://catboy.best}.} A beatmapset holds one audio recording together with charts at several difficulties, and the same song can appear under more than one set, such as a TV-size cut beside the full song. We group songs by normalized title and audio checksum, and split 80/10/10 by song. A song that appears in more than one mode is placed in the same split in each. For \textit{Clone Hero} we use a corpus of charts paired with audio from Chorus Encore,\footnote{Retrieved from \url{https://www.enchor.us}.} keeping the Expert lead guitar part of each chart and the corpus's own 80/10/10 split. We draw 210.6 hours, as much as \textit{taiko} and \textit{standard} together. Its labels carry no well-defined difficulty rating, so we condition on note density instead. In one joint model it takes the place of those two modes. Table~\ref{tab:data} summarizes the datasets.

\begin{table}[!t]
\centering
\footnotesize
\setlength{\tabcolsep}{4pt}
\begin{tabular}{lrrrrl}
\toprule
Mode & Songs & Charts & Charts/song & Hours & Stars \\
\midrule
\textit{standard} & 2{,}111 & 8{,}687 & 4.12 & 117.9 & 1.03--13.15 \\
\textit{taiko}    & 1{,}845 & 7{,}734 & 4.19 &  92.6 & 0.63--10.36 \\
\textit{mania}    & 1{,}955 & 6{,}268 & 3.21 &  99.5 & 0.59--10.07 \\
\midrule
\textit{Clone Hero} & 2{,}782 & 2{,}782 & 1.00 & 210.6 & --- \\
\bottomrule
\end{tabular}
\caption{Dataset statistics.
}
\label{tab:data}
\end{table}
\subsection{Training}
\label{ssec:train}

We train with a next-token cross-entropy objective, using Adam with a batch size of 64, a learning rate of $3\times10^{-4}$ and weight decay $10^{-4}$. The learning rate follows a cosine decay to zero after a linear warm-up over 2\% of training.
We run 800,000 steps on \textit{standard} and on the joint models, and 450,000 on \textit{taiko}. The joint model draws each mode in proportion to its training samples, which gives \textit{mania} 26\% of the draws, and the model trained with \textit{Clone Hero} keeps this share. We train two \textit{mania}-only models, one for the matching 208,544 steps and one for the full 800,000. As augmentation, we slice each song at five offsets 1\,s apart and mix them within a batch. We draw samples with inverse-frequency weights over star-rating bands (note density for \textit{Clone Hero}) to balance rare and common difficulties. We keep the checkpoint with the lowest validation loss, taken over all modes for the joint models.

\begin{table*}[!t]
\centering
\footnotesize
\setlength{\tabcolsep}{5pt}
\begin{tabular}{lrrrrrrr@{\hskip 16pt}rrrrrrr}
\toprule
& \multicolumn{7}{c}{\textit{standard}} & \multicolumn{7}{c}{\textit{taiko}} \\
\cmidrule(lr){2-8}\cmidrule(l){9-15}
& & & \multicolumn{3}{c}{Onset} & \multicolumn{2}{c}{Event}
& & & \multicolumn{3}{c}{Onset} & \multicolumn{2}{c}{Event} \\
\cmidrule(lr){4-6}\cmidrule(lr){7-8}\cmidrule(lr){11-13}\cmidrule(l){14-15}
Encoder & Trained & NLL$\downarrow$ & P & R & F1 & F1 & Type
        & Trained & NLL$\downarrow$ & P & R & F1 & F1 & Type \\
\midrule
C-LSTM~\cite{ddc}             & 0.7M   & 1.388 & 0.628 & \textbf{0.688} & 0.654 & 0.420 & 0.637
                              & 0.7M   & 0.894 & 0.753 & 0.741 & 0.743 & 0.391 & 0.524 \\
Transformer~\cite{muscriptor} & 240.2M & 1.447 & 0.507 & 0.587 & 0.541 & 0.323 & 0.591
                              & 239.9M & 0.948 & 0.706 & 0.715 & 0.707 & 0.367 & 0.516 \\
\midrule
YourMT3+~\cite{yourmt3}       & 1.6M   & 1.429 & 0.600 & 0.642 & 0.617 & 0.388 & 0.622
                              & 0.8M   & 0.944 & 0.713 & 0.728 & 0.716 & 0.370 & 0.514 \\
MuScriptor~\cite{muscriptor}  & 0.5M   & \textbf{1.359} & \textbf{0.646} & 0.675 & \textbf{0.657} & \textbf{0.433} & \textbf{0.653}
                              & 0.3M   & \textbf{0.846} & \textbf{0.754} & 0.747 & 0.746 & \textbf{0.403} & \textbf{0.538} \\
MERT~\cite{mert}              & 0.5M   & 1.369 & 0.645 & 0.664 & 0.651 & 0.420 & 0.639
                              & 0.3M   & 0.848 & 0.748 & \textbf{0.762} & \textbf{0.751} & 0.402 & 0.534 \\
\midrule
Human\_adj.                  & ---    & ---   & ---   & ---   & 0.752 & 0.549 & 0.721
                              & ---    & ---   & ---   & ---   & 0.796 & 0.580 & 0.725 \\
Human\_any                   & ---    & ---   & ---   & ---   & 0.673 & 0.456 & 0.660
                              & ---    & ---   & ---   & ---   & 0.657 & 0.451 & 0.672 \\
\bottomrule
\end{tabular}
\caption{Onsets are matched at 25\,ms; Event F1 also requires the note type, and Type is the type accuracy over matched onsets. NLL is the teacher-forced cross-entropy per token; other scores are averaged over charts. Trained is what is learned between the audio and the decoder --- encoder and projection when trained from scratch, projection alone when frozen --- and excludes the decoder (25M \textit{standard}, 4M \textit{taiko}).
}
\label{tab:main}
\end{table*}

\subsection{Evaluation}
\label{ssec:eval}

For each reference chart in the test split we generate one chart, conditioning on its star rating. At inference we also take its timing points, the tempo and beat positions stored with the beatmap. They are not a model input and leave note times unchanged: they only convert slider lengths from beats into time, which keeps the comparison independent of beat tracking. We use nucleus sampling, at temperature 1.1 and $p=0.9$ for \textit{standard} and at 1.0 and 0.8 for \textit{taiko} and \textit{mania}, values chosen in preliminary runs; \textit{standard} is sampled more broadly because the same rhythm admits many valid placements.

Onsets are compared within a tolerance window, as in previous chart generation~\cite{ddc,yi2023}. In \textit{mania}, simultaneous notes count as separate onsets. We match onsets one to one in time order, each generated onset taking the earliest unmatched reference onset; this yields as many matched onsets as the maximum matching standard in transcription~\cite{mireval}. We report onset precision, recall and F1 at 25\,ms, and at the same tolerance event F1, which requires the note type to match, and type accuracy over the matched pairs. The type is the one each mode encodes (Sec.~\ref{sec:chart}), so event F1 does not score the column in \textit{mania} or the position in \textit{standard}. Since many charts are valid for one song, we also report the negative log-likelihood (NLL) the model assigns to the reference chart, which covers these as well.

The same metrics score pairs of human charts of one song, where the easier chart stands in for the generated one. We use a pair at adjacent star ratings (\textit{Human\_adj.}), the closest the data offers to a model conditioned on one rating, and a pair of any two charts of the song (\textit{Human\_any}).

\section{Results}
\label{sec:results}

\subsection{Transfer from music transcription}
\label{ssec:results1}

As shown in Table~\ref{tab:main}, MuScriptor outperforms the C-LSTM on NLL, event F1 and type accuracy in both modes and is comparable on onset F1: the frozen transcription encoder fits human charts more closely than an encoder designed and trained for the task. It also achieves a lower NLL than the same architecture trained from scratch, and a higher event F1 (0.433 against 0.323). Since those two take the same input and differ only in the encoder weights, the gain comes from pre-training rather than the architecture.

If the gain comes from the pre-training task, a stronger transcription model should give a stronger chart model. MuScriptor reports the higher frame-level transcription results of the two AMT encoders~\cite{muscriptor}, and leads YourMT3+ on every score in both modes, consistent with transcription quality carrying over to chart generation, to the object types as well as the onsets.

MERT, pre-trained without transcription labels, also outperforms the Transformer trained from scratch on every score and is comparable to MuScriptor on onset F1, but its type accuracy is lower in both modes. Pre-training alone thus accounts for most of the gain over training from scratch, and the transcription objective slightly improves the object placed at each onset.

\subsection{Transfer across rhythm games}
\label{ssec:results2}

We measure cross-mode transfer on \textit{mania}, the mode with the fewest charts (Table~\ref{tab:data}). As shown in Table~\ref{tab:mania}, the joint model outperforms a \textit{mania}-only model on onset F1 (0.738 against 0.728) and event F1 (0.506 against 0.492). The gain does not come from a longer schedule: the single-mode model scores the same when trained for the full 800,000 steps (0.728 onset F1, 0.491 event F1). The \textit{mania}-only model has the lower NLL (0.473 against 0.501), as the joint model's checkpoint is selected on the validation loss over all modes.

The improvement falls on the object the three modes share. The tap rises from 0.599 to 0.612, while the two long-note types fall in exchange.
Training \textit{mania} with \textit{Clone Hero} instead lowers both scores to 0.707 onset F1 and 0.482 event F1, below the \textit{mania}-only model as well. \textit{Clone Hero} is lane-based like \textit{mania}, unlike the other osu! modes, so similar objects alone do not bring the gain.
We speculate that two differences between the games account for this. A \textit{Clone Hero} chart follows the melody of a single instrument part, most often the guitar, whereas the osu! modes follow the beat of the whole song and move between instruments. The two games thus chart the same audio under different conventions. Consistent with this, the decrease falls on the onsets and not on the types, as the ratio of event F1 to onset F1 stays between 0.68 and 0.69 across the three models. The two games also draw on different music. Metal and rock account for 69\% of our \textit{Clone Hero} songs, and electronic, anime and video game music for 70\% of our \textit{mania} songs. Shared objects would then help only when the games also agree on what a chart follows and on the music they cover.

\begin{table}[!t]
\centering
\footnotesize
\setlength{\tabcolsep}{3.5pt}
\begin{tabular}{lrrrrrr}
\toprule
& & & & & \multicolumn{2}{c}{Hold} \\
\cmidrule(l){6-7}
Trained on & NLL$\downarrow$ & Onset & Event & Tap & Start & End \\
\midrule
\textit{mania} only & \textbf{0.473} & 0.728 & 0.492 & 0.599 & \textbf{0.122} & \textbf{0.072} \\
\textit{mania} + \textit{standard} + \textit{taiko} & 0.501 & \textbf{0.738} & \textbf{0.506} & \textbf{0.612} & 0.099 & 0.056 \\
\textit{mania} + \textit{Clone Hero} & 0.506 & 0.707 & 0.482 & 0.584 & 0.096 & 0.054 \\
\bottomrule
\end{tabular}
\caption{The MuScriptor encoder and the decoder are fixed across rows; only the training data varies.
Onset F1 at 25\,ms; the last three columns are per-class F1.}
\label{tab:mania}
\end{table}

\section{Conclusion}
\label{conclusion}

We studied transfer learning for chart generation along two dimensions. For transfer from a pre-trained audio encoder, the frozen MuScriptor encoder outperforms the same architecture trained from scratch and fits human charts more closely than the C-LSTM of DDC. The stronger of the two transcription models gives the stronger chart model. Most of the gain over training from scratch comes from pre-training, and the transcription objective slightly improves the object placed at each onset.
For transfer across rhythm games, training one model on three osu! modes raises event F1 on \textit{mania}, the mode with the fewest charts. Training with \textit{Clone Hero} instead lowers it, which suggests that shared objects help only when the games agree on charting style and on the music they cover.

For future work, we aim to take both dimensions of transfer further. Transfer from AMT could run in the reverse direction, since charts mark note onsets and could serve as training data for an AMT model. For transfer across rhythm games, games whose charts follow the same part of the music, such as the osu! modes and the drum charts of \textit{Clone Hero}, could share their charts. Pre-training one model on charts from many games would test this.

\vfill\pagebreak

\section{Acknowledgments}
\label{sec:ack}

This work was supported by the Ministry of Education of the Republic of Korea and the National Research Foundation of Korea (NRF-2024S1A5C3A03046168) and by funding from Sony AI.

\bibliographystyle{IEEEbib}
\bibliography{strings,refs}

\end{document}